\documentclass[12pt, draftclsnofoot, onecolumn,comsoc]{IEEEtran}

\usepackage{booktabs} 
\usepackage{graphicx}
\graphicspath{{./Bilder/}}

\usepackage{mathtools}
\usepackage{amssymb}
\usepackage[cmintegrals]{newtxmath}
\usepackage{chemformula}
\usepackage{bm}
\usepackage{comment}
\usepackage{siunitx}
\usepackage{epstopdf}
\usepackage[process=auto]{pstool}
\usepackage{algorithm}
\usepackage{algorithmic}
\usepackage{multirow}
\newtheorem{remk}{Remark}
\usepackage{siunitx}
\usepackage{cite}

\newcommand{\pumprate}{\rho}
\newcommand{\ceqdark}{c^{\textrm{eq}}_{\textrm{in}}}

\newcommand{\cinf}{c^{\infty}}
\newcommand{\ddt}{\frac{\mathrm{d}}{\mathrm{d}t}}

\DeclareSIUnit\rpm{rpm}

\begin{document}
\title{Biological Optical-to-Chemical Signal Conversion Interface: A Small-scale Modulator for Molecular Communications}

\author{%
Laura~Grebenstein,
Jens~Kirchner,
Renata~Stavracakis~Peixoto,
Wiebke~Zimmermann,
Florian~Irnstorfer,
Wayan~Wicke,~\IEEEmembership{Student Member,~IEEE,}
Arman~Ahmadzadeh,~\IEEEmembership{Student Member,~IEEE,}
Vahid~Jamali,~\IEEEmembership{Student Member,~IEEE,}
Georg~Fischer,~\IEEEmembership{Senior Member,~IEEE,}
Robert~Weigel,~\IEEEmembership{Fellow,~IEEE,}
Andreas~Burkovski,
and Robert~Schober,~\IEEEmembership{Fellow,~IEEE}\\
\thanks{This work was supported in part by the German Research
Foundation (Project SCHO 831/7-1), the Friedrich-Alexander University 
Erlangen-N\"urnberg (FAU) under the Emerging Fields Initiative (EFI), and the STAEDTLER Foundation.
This paper has been accepted for presentation in part at ACM/IEEE NanoCom 2018 \cite{Grebenstein_Biological_2018}.
\textit{(Laura Grebenstein and Jens Kirchner are co-first authors.) (Corresponding author: Wayan Wicke.)}}%
\thanks{L.~Grebenstein, R.~Stavracakis~Peixoto, and A.~Burkovski are with the Institute for Microbiology at the Friedrich-Alexander University Erlangen-N\"urnberg (FAU) (e-mail: laura.grebenstein@fau.de).}%
\thanks{J.~Kirchner, W.~Zimmermann, G.~Fischer, and R.~Weigel are with the Institute for Electronics Engineering at FAU (e-mail: jens.kirchner@fau.de).}%
\thanks{F.~Irnstorfer, W.~Wicke, A.~Ahmadzadeh, V.~Jamali, and R.~Schober are with the Institute for Digital Communications at FAU (e-mail: wayan.wicke@fau.de; arman.ahmadzadeh@fau.de; vahid.jamali@fau.de; robert.schober@fau.de).}%
\vspace{-1.5cm}}

\maketitle

\begin{abstract}
Although many exciting applications of molecular communication (MC) systems are envisioned to be at microscale, the MC testbeds  reported in the literature so far are mostly at macroscale.
This may partially be due to the fact that controlling an MC system at microscale is challenging.
To link the macroworld to the microworld, we propose and demonstrate a biological signal conversion interface that can also be  seen as a microscale modulator.
In particular, the proposed interface transduces an optical signal, which is controlled using a light-emitting diode (LED), into a chemical signal by changing the pH of the environment.
The modulator is realized using \textit{Escherichia coli} bacteria as microscale entity expressing the light-driven proton pump gloeorhodopsin from \textit{Gloeobacter violaceus}.
Upon inducing external light stimuli, these bacteria locally change their surrounding pH level by exporting protons into the environment.
To verify the effectiveness of the proposed optical-to-chemical signal converter, we analyze the pH signal measured by a pH sensor, which serves as receiver.
We develop an analytical parametric model for the induced chemical signal as a function of the applied optical signal.
Using this model, we derive a training-based channel estimator which estimates the parameters of the proposed model to fit the measurement data based on a least square error approach.
We further derive the optimal maximum likelihood detector and a suboptimal low-complexity detector to recover the transmitted data from the measured received signal.
It is shown that the proposed parametric model is in good agreement with the measurement data.
Moreover, for an example scenario, we show that the proposed setup is able to successfully convert an optical signal representing a sequence of binary symbols  into a chemical signal with a bit rate of \SI{1}{bit/\minute} and recover the transmitted data from the chemical signal using the proposed estimation and detection~schemes.
The proposed modulator may form the basis for future MC testbeds and applications at microscale.
\end{abstract}


\section{Introduction}

Molecular communication (MC) systems embed information into the characteristics of signaling molecules. This is very different from  conventional electromagnetic- (EM-) based  communication systems that embed data into the properties of EM waves \cite{Nakano_Molecular_2013,Farsad_comprehensive_2016}. MC systems are suitable for communication at small scale and in fluids where  EM-based communication is inefficient or even infeasible. Functioning MC systems are envisioned to enable new revolutionary applications including sensing of a target substance  in 
biotechnology, targeted drug delivery in medicine, and  monitoring of oil pipelines or chemical reactors in industrial applications~\cite{Farsad_comprehensive_2016}.

An important step towards realizing the aforementioned applications is the development of testbeds that allow the verification of the theoretical channel models and the transmission strategies proposed in the MC literature. To this end, MC testbeds based on spraying alcohol  into open space and  using acids and bases within closed vessels have been developed in \cite{Farsad_Tabletop_2013} and \cite{Farsad_Novel_2017}, respectively. These testbeds have been extended to multiple-input multiple-output (MIMO) systems \cite{Koo_Molecular_2016}, and improved channel models have been proposed to account for discrepancies between theory and experimental results \cite{Farsad_Channel_2014}. Recently,  an in-vessel MC testbed was proposed in \cite{TestBed_Harold} that uses specifically
designed magnetic nanoparticles as information carriers. These nanoparticles are biocompatible, clinically
safe, and do not interfere with chemical processes like alcohol \cite{Farsad_Tabletop_2013} or acids and bases \cite{Farsad_Novel_2017}  
may do. Nevertheless, the aforementioned MC testbeds are all at macroscale, i.e., their dimensions are on the order of several tens of centimeters, whereas many prospective applications of MC systems are envisioned to be at microscale. 

Biologically inspired experimental studies related to MC have been conducted in \cite{Krishnaswamy_Time_2013,Felicetti_Modeling_2014,Nakano_Microplatform_2008,Nakano_Interface_2014,Akyildiz_testbed_2015}. In particular, in  \cite{Krishnaswamy_Time_2013}, bacterial populations were used as transceivers connected through a microfluidic channel.  In \cite{Felicetti_Modeling_2014}, soluble CD40L molecules were released from platelets (as transmitter) into a fluid medium that upon contact triggered the activation of endothelial cells (as receiver). Moreover, in \cite{Nakano_Microplatform_2008}, a microplatform was designed  to demonstrate the propagation of molecular signals through a line of patterned HeLa cells (human cervical cancer cells) expressing gap junction channels. In \cite{Nakano_Interface_2014}, artificially synthesized materials were embedded into the cytosol of living cells and, in response to stimuli induced in the cells, emitted fluorescence that could be externally detected by fluorescence microscopy. Similarly, in \cite{Akyildiz_testbed_2015}, the response of genetically engineered \textit{Escherichia coli} (\textit{E. coli}) bacteria to the surrounding molecules was used as the basis for the design of a biological receiver. We note that the systems in \cite{Felicetti_Modeling_2014,Nakano_Microplatform_2008,Nakano_Interface_2014,Akyildiz_testbed_2015} were demonstrated for a single shot transmission. Furthermore, the setup with continuous transmission in \cite{Krishnaswamy_Time_2013} achieves low data rates on the order of one \si{bit/\hour}. 

One particular challenge for designing and operating microscale MC testbeds are the difficulties associated with  controlling an MC system at microscale. To address this issue, in this paper, we propose a biological signal conversion interface  which converts an optical signal, which can be easily controlled using a light-emitting diode (LED), into a chemical signal in form of a change of the pH of the environment. This setup can be  seen as a microscale modulator\footnote{Throughout the paper, we use the terms ``optical-to-chemical signal converter'' and ``modulator'' interchangeably.}. The modulator is realized using \textit{E. coli} bacteria that express the light-driven proton pump gloeorhodopsin (GR), a bacterial type~I rhodopsin. Upon inducing external light stimuli, these bacteria change their surrounding pH level by exporting protons into the environment. The authors of \cite{Choi_Cyanobacterial_2014} examined the proton flux due to illumination of \textit{E. coli} bacteria expressing GR but not for application in  MC systems. In particular, one proton can be transferred to the periplasmic space in less than \SI{1}{\milli\second} from an almost inexhaustible pool of protons inside the cell arising from the cell's energy metabolism \cite{Lanyi_Proton_2006}. As a result, in a bacterial suspension, the change of proton concentration in the surrounding medium can be detected within a few seconds as a change of pH. The bacteria in the proposed system automatically regenerate the surrounding pH level by back pumping mechanisms \cite{Capaldi_ATPase_2000,HOSKING_2006,ZHANG_2014}. Nevertheless, since the kinetics of the regeneration are much slower than the proton efflux rates, we expect that the considered system  achieves a relatively fast signal conversion in comparison with the setup in \cite{Krishnaswamy_Time_2013} where a chemical signal was generated based on gene expression. In particular, compared to the data rates on the order of one \si{bit/\hour}  reported in \cite{Krishnaswamy_Time_2013}, the proposed testbed achieves significantly higher data rates on the order of one \si{bit/\minute}. 

To test the effectiveness of the proposed optical-to-chemical signal converter, we analyze the pH signal measured by a pH sensor, which serves as the receiver. We develop an analytical parametric model for the induced chemical signal as a function of the applied optical signal. Using this model, we derive a training-based parametric channel estimator which, using the least square error as estimation performance criterion, estimates the model parameters based on known pilot symbols inserted at the beginning of each transmission frame. In addition, we derive the optimal maximum likelihood~(ML) detector and a suboptimal low-complexity detector to recover the transmitted data from the measured received signal. Furthermore, adaptive transmission strategies are proposed to deal with the time-varying nature of the considered biological system. It is shown that the proposed parametric model is in good agreement with the measurement data. Moreover, for an example scenario, we show that the proposed setup is able to successfully convert an optical signal representing a sequence of binary symbols  into a chemical signal for a bit rate of \SI{1}{bit/\minute}  and recover the transmitted data from the chemical signal using the proposed estimation and detection~schemes. We note that the proposed setup can serve as the basis for the development of testbeds using other light-driven pumps that generate other chemical signals, e.g., Na$^+$ and K$^+$ ions~\cite{LightPump_2013,LightPump_2017}.

The rest of the paper is organized as follows. The proposed system is presented in Section~\ref{Sec:SysMod}, and the testbed setup is described in detail in Section~\ref{Sec:TestBed}. The proposed parametric channel model is developed in Section~\ref{Sec:Analytic}, and the parametric channel estimator and the optimal and suboptimal detectors are derived in Section~\ref{Sec:Comm}. Finally, measurement results are reported in Section~\ref{Sec:Results}, and the paper is concluded in Section~\ref{Sec:Conclusion}. 

\section{System Setup and Preliminaries}\label{Sec:SysMod}
In this section, first an overview of the experimental system is provided. Subsequently, the photocycle of bacteriorhodopsin, the main  biological mechanism that is exploited for the proposed microscale modulator, is discussed in  detail. Finally, we discuss the biological kinetics of the bacterial system which will be used to derive the analytical model in Section~\ref{Sec:Analytic}.

\begin{figure}[t]
\center
    \includegraphics[width=0.6\columnwidth]{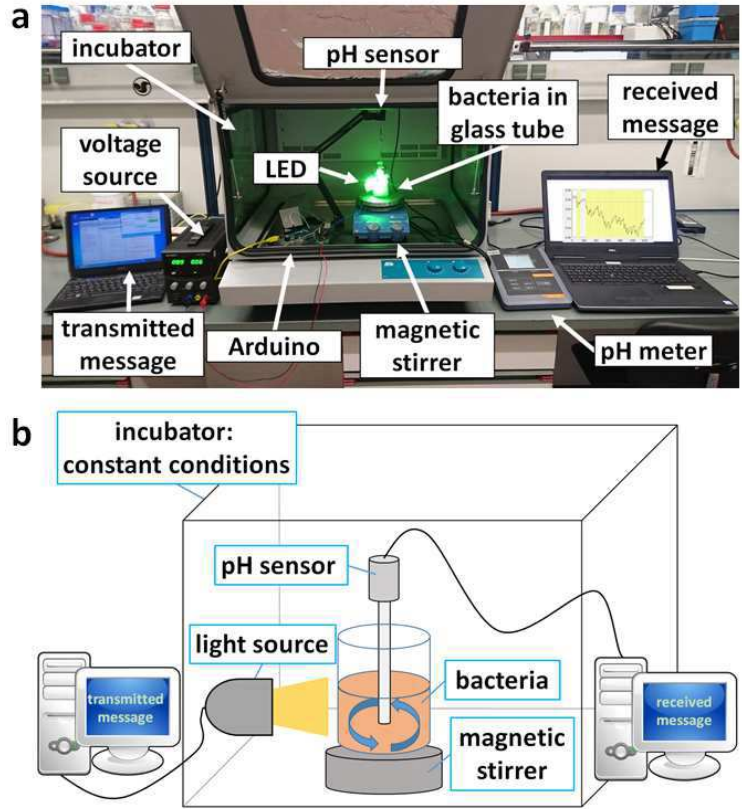}
    \caption{%
        Biological modulator model. (a) Benchtop experimental setup; (b) Schematic illustration.
    }
    \label{Fig:SysMod}
\end{figure}

\subsection{System Overview}
\begin{figure}[t]
\center
    \includegraphics[width=0.6\columnwidth]{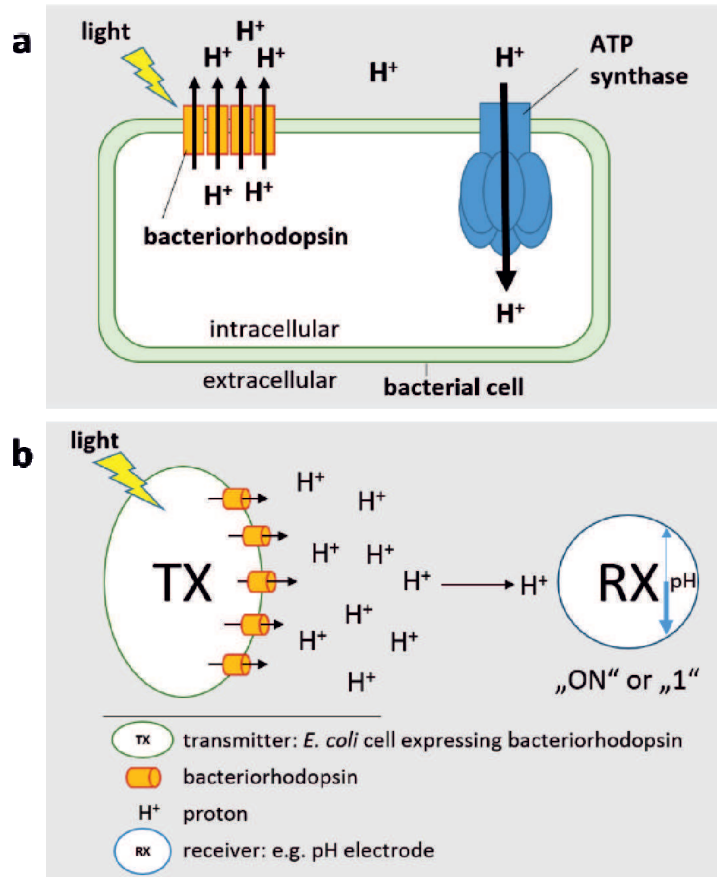}
    \caption{%
        The light-driven proton pump bacteriorhodopsin.
        (a) Biological function of bacteriorhodopsin in a native cell;
        (b) Schematic transmission model.
    }
    \label{Fig:Bacteria}
\end{figure}

The developed testbed is shown in Fig.~\ref{Fig:SysMod}a and schematically illustrated in Fig.~\ref{Fig:SysMod}b. The proposed modulator is based on \textit{E. coli} bacteria expressing bacteriorhodopsin in their cell membrane for easy-to-control optical signal conversion. A glass tube containing the bacterial suspension is installed in a light-isolated incubator in order to keep environmental conditions, such as temperature, constant. An LED is focused on the  bacterial suspension and is controlled using an Arduino microcontroller and a personal computer (PC). Thereby, the information generated by the transmitter PC is encoded into an optical signal using on-off keying (OOK) modulation. The optical signal is then converted by the bacteria to a chemical signal in form of a pH change. In particular, upon illumination the bacteriorhodopsin in the bacteria plasma membrane pump protons out into the surrounding medium, see Fig.~\ref{Fig:Bacteria}. In an aequeous solution, the proton pumping reduces the pH according to $\text{pH}=-\log_{10}(c_{\text{H}^+})$ where $c_{\text{H}^+}$ is the concentration of the protons in \si{\mol/\liter} \cite{Farsad_Novel_2017}. In order to evaluate the efficiency of the proposed signal converter, we deploy a pH sensor in the bacterial suspension which tracks the pH variations over time. This sensor reports the pH values to a receiver PC for signal processing. The technical details of the components of the testbed and the cultivation of the bacteria are provided in Section~\ref{Sec:TestBed} and the modulation and detection schemes used to collect and process the measurement data are presented in Section~\ref{Sec:Comm}.  

\subsection{Bacteriorhodopsin Photocycle}

The modulator in this testbed consists of bio-engineered bacteria. The cells carry a vector
DNA bearing the gene encoding GR\footnote{GR is a specific bacteriorhodopsin belonging to the family of bacterial type~I rhodopsins. Throughout this paper, we use ``GR" and ``bacteriorhodopsin" interchangeably.}. After chemically induced gene expression the protein is inserted into the plasma membrane of the cell, which is not permeable for charged ions like protons. The GR protein provides a gate through the membrane via seven transmembrane domains formed by amino acid helices. Due to a hydrophobic barrier on the cytoplasmic side, the protein does not provide any transport of molecules in the ground state. To perform proton-transfer, a chromophore group, the all-trans retinal, is needed. The retinal is integrated into the protein and acts as a biochemical pumping lever. The photocycle of bacteriorhodopsin was investigated extensively over the past decades in the biology literature \cite{Lanyi_Bacteriorhodopsin_2004}. In the ground state, retinal is in the all-trans configuration and a proton is bound to the residue of amino acid Asp96 of GR on the cytoplasmic side. By the energy of one photon, the retinal is subject to a trans$\rightarrow$cis transition at carbon atom C14 and thereby performs the lever action. As a result, one proton is transferred from the Schiff base to the residue of amino acid Asp85 on the periplasmic side of the protein within picoseconds. Investigations of the bacteriorhodopsin photocycle strongly suggest that the protonation of Asp85 causes the passage of a proton through a water network embedded in the amino acid residues of the protein on the extracellular side \cite{Patterson_Ultrafast_2010}. Hence, the proton can move through the plasma membrane against the electrochemical potential along the amino acid residues inside the protein. Furthermore, Asp85 reprotonates the Schiff base and the retinal regenerates to the ground state, ready for a new cycle.

The photo-isomerization of the retinal and the release of one proton to the extracellular space is the fastest known bacterial photoreaction and is performed in less than \SI{1}{\milli\second} \cite{Patterson_Ultrafast_2010}. However, the regeneration from the excited state to the ground state takes \SI{15}{\milli\second} which makes it the time-limiting factor in the photocycle. By increasing the proton-gradient, in the natural host, the polarization of the membrane is  used to drive, e.g., an ATP synthase to convert light energy to chemical energy, see Fig.~\ref{Fig:Bacteria}a, in the respiratory chain or to drive the flagellar apparatus of the bacteria. 

\begin{remk}
It is noted that light is one of the most important external signals used to convey information from the external world to biological systems. In fact, in addition to the ion transporting rhodopsins, there are also sensory rhodopsins functioning as light-signal transducers in nature \cite{Kaneko_Conversion_2017}. Therefore, natural organisms make use of light not only as energy source but also as information signal, similar to the proposed modulator. 
\end{remk}

\subsection{Biological Kinetics of the Bacterial System}\label{Sec:Kinetic}
The biological membrane forms a compartment border, which allows to generate electrical and chemical gradients. In a steady ground state the proton concentration is in chemo-osmotical balance and will not change due to the impermeability of the plasma membrane. Only energy surplus can provide an active export of protons against the natural gradient. Proteins like GR can transform the energy of, for example, one photon and use it for active transport processes. Other proteins providing active transport are gated by voltage, pressure, signaling molecules or by using other ion gradients for co-transport. The kinetics of these proteins depend on the activation energy, the properties of the molecule they transport, and the gradient across the membrane. In the proposed system, the bacterial cells carry a number of proteins with different functions inside the cell and in the membrane. With the activation of proton efflux by the energy of light, the gradient changes fast according to the GR kinetics (photocycle; see also Section~II.B). With the increasing gradient, the energy needed to pump protons against the gradient also increases. In a measured pH signal, only net fluxes of protons can be detected. Besides the changing kinetics of GR, other proteins are active in parallel, driven by the proton gradient. By this, a net proton efflux can be observed, comprising more exported protons by GR than imported protons by other proteins, like ATPase. With the rising proton gradient, saturation will be reached when the net flux is zero. By turning off the light, the active transport against the gradient immediately stops and only proton influx can be observed, depending on the kinetics of the importing proteins. Since there is no external energy surplus and the influx is only driven by the chemoosmotical gradient, the regeneration kinetics in the dark differ strongly from the kinetics in the illumination phase. The influx can be observed as a pH increase, until saturation of the ground state is reached again. 

\section{Detailed Description of the Experimental Setup}\label{Sec:TestBed}
In this section, we first describe the procedures for cultivation of the bacteria, formation of the spheroplasts (bacterial cells where the outer membrane is removed) that are needed for efficient proton pumping, and taking measurements. Subsequently, we provide a brief discussion on the variability that is expected to occur in MC systems that employ biological~components. 

\subsection{Bacterial Cultivation}

In this paper, we use genetically modified \textit{E. coli} bacteria, namely the strain \textit{E. coli} $\textrm{DH5}\alpha\textrm{Mcr}$, carrying the vector DNA pKJ900 with the gene encoding GR from \textit{Gloeobacter violaceus} under control of the chemically induced \textit{ptac} promoter that was proposed in \cite{Choi_Cyanobacterial_2014}.
Bacteria from a dry agar culture were pre-cultured for \SI{6}{\hour} at \SI{37}{\celsius} and shaked at \num{200} revolutions per minute (\si{\rpm}) in \SI{20}{\milli\liter} lysogeny broth (LB) medium (i.e., \SI{10}{\gram/\liter} tryptone, \SI{10}{\gram/\liter} \ch{NaCl}, \SI{5}{\gram/\liter} yeast extract)  with \SI{25}{\micro\gram/\milli\liter} chloramphenicol, to select bacteria with antibiotic resistance genetically encoded in the vector DNA. Subsequently, for the main culture, \SI{400}{\milli\liter} LB with chloramphenicol was inoculated to a final optical density at \SI{600}{\nano\meter} of OD\textsubscript{\SI{600}{\nano\meter}}$=$\numrange{0.01}{0.02} (approximately $0.02\times (8\times 10^{8})=1.6\times 10^{7}$ \si{cells/\milli\liter}), and incubated for \SI{1}{\hour} at \SI{37}{\celsius} at \SI{175}{\rpm} constant shaking, to adapt to the fresh medium conditions. Thereafter, \SI{100}{\micro\mol/\liter} isopropyl-$\beta$-D-thiogalactopyranosid (IPTG) for chemical induction of the transcription, and \SI{10}{\micro\mol/\liter} retinal were added. Since \textit{E. coli} cells do not produce retinal, it has to be supplied by the medium in which the cells grow. For the success of the later performed spheroblast formation, the amount of expressed and retinal inbound GR was crucial. A one week old (but not older) solution of retinal showed better results in terms of chemical signal strength compared to a fresh solution. Afterwards, the main culture was incubated at \SI{35}{\celsius} and \SI{50}{\rpm} in the dark.

\subsection{Spheroplast Formation}
Since GR is located in the plasma membrane, protons are pumped to the periplasmic space between the cytosolic and the outer membrane (OM). Therefore, most released protons are trapped and cannot easily reach the extracellular environment. To address this issue, we standardized a protocol based on sonication with the aim to remove the OM so that the protons are released directly into the surrounding medium. Among many protocols already described in the literature, using lysozyme resulted in a high and pure yield of spheroplasts but in a lower final volume and concentration of cells compared to sonication \cite{Hobb_Evaluation_2009,Liu_effect_2006,Marvin_highly_1987,Mohan_Autolytic_1965}. Thus, OM removal by sonication is the method that better fitted the requirements of the testbed. The IPTG-induced cells were harvested after 16~h of incubation by centrifugation ($4000g$, \SI{5}{\minute}, \SI{4}{\celsius}). GR binding retinal causes a typical red color of the cells, which was also used as a preliminary tool to optically evaluate the cells for the success of the spheroblast formation. Afterwards, the cells were resuspended in \SI{320}{\milli\liter} \SI{0.9}{\percent} \ch{NaCl} in total and exposed to 6 times of \SI{20}{\second} sonication bath with ice (\SI{10}{\percent} power in Bandelin Sonorex Digital 10P) and \SI{20}{\second} regeneration.  After centrifugation with $8000g$ for \SI{10}{\minute}, \SI{4}{\celsius}, the cell pellet was resuspended again in \SI{320}{\milli\liter} \SI{0.9}{\percent} \ch{NaCl}. The proportion of OM removal is strongly dependent on the dilution during sonication. In total, 6 cycles of sonication were performed. After the last centrifugation, the spheroplasts were resuspended in an unbuffered, osmotically balancing solution (\SI{120}{\milli\mol/\liter} \ch{NaCl}, \SI{10}{\milli\mol/\liter} \ch{MgCl2}, \SI{10}{\milli\mol/\liter} \ch{KCl}, \SI{10}{\milli\mol/\liter} \ch{MgSO4}, \SI{100}{\micro\mol/\liter} \ch{CaCl2}) with a pH of \num{5.5} and adjusted to an OD\textsubscript{600~nm} of 15 (approximately $15\times (8\times 10^{8})=\SI{1.2e10}{cells/\milli\liter}$). The resulting solution was a mixture of spheroplasts (\SIrange{50}{60}{\percent}, optically estimated), cells with partly removed OM, and intact cells. In a reaction tube, \SI{6}{\milli\liter} of the cell suspension were incubated for \SI{5}{\hour} in \SI{35}{\celsius} in the testbed setup, see Fig.~\ref{Fig:SysMod}, stirred at level 6 of an IKA\textsuperscript{\textregistered} RCT basic magnetic stirrer, and finally dark adapted to stable osmolaric and pH conditions before it was used for signal generation.

\subsection{Measurement}
The bacteria, constantly incubated in a dark environment, were illuminated by an LED with optical power \SI{1}{\watt}, which operated at wavelength \SI{550}{\nano\meter} due to the maximum absorption of GR \cite{Choi_Cyanobacterial_2014}. The LED was controlled by a custom Matlab\textsuperscript{\textregistered} (MathWorks\textsuperscript{\textregistered}, Natick, MA, United States) graphical user interface (GUI), which allowed mapping a user-defined bit sequence to an appropriate sequence of light stimuli. The transmitter PC was connected to an Arduino Mega 2560 (Rev. 3) microcontroller via serial connection. The GUI controled one of the digital output pins of the microcontroller, which in turn provided the control signal for the custom LED driver circuit PT4115 (CR Powtech, Shanghai, China). The measurement was performed when the temperature was stable at $35\pm\SI{0.2}{\celsius}$ and the pH was adapted to between \numrange{5.6}{5.8}, since this was the most effective operating range to generate a strong signal from the bacteria \cite{Wang_Spectroscopic_2003}. The pH signal in general was documented for at least \SI{30}{\minute} to ensure stability. The absolute pH level was detected with a SenTix 950 (Xylem Analytics, WTW, Weilheim, Germany) microelectrode using the potentiometric pH meter inoLab\textsuperscript{\textregistered} Multi 9310 IDS (Xylem Analytics, WTW, Weilheim, Germany). Since our main objective was to characterize the optical-to-chemical signal conversion, the pH microelectrode was inserted directly into the bacterial solution. The measured real-time data were continuously streamed via serial connection to the receiver PC, where they were analyzed, displayed, and stored by a custom Matlab\textsuperscript{\textregistered} GUI.

\subsection{Variability in Biological Systems}\label{Sec:Variability}

The proposed testbed is based on living biological organisms. Hence, we expect conditional unique characteristics that are usually not observed in synthetic non-living systems. In particular, in the following, we highlight the factors that may cause variations in the overall system response and may help in interpreting the experimental data. These factors include, but are not limited to, the portion of spheroplasts in the cell mixture and the number of GR molecules with integrated retinal in each cell. Moreover, as bacteria age, changes in the system response could also arise from degenerating processes in the cell. These factors may lead to a baseline drift in the chemical signal over time, cf. Section~\ref{Sec:Results}. To minimize these effects, we followed a careful protocol for preparation of the bacteria and the  measurement procedure as discussed in the previous subsections. Nevertheless, residual variations still exist that will be studied and modeled in the next section. Determining for example the exact numbers of spheroplasts and functional GR molecules, and the development of efficient protocols to reduce or eliminate variations in these numbers, are important topics for future research. One option in this regard will be to use artificial
membrane vesicles, i.e., liposomes. In such systems only GR of a defined concentration can be
embedded in the modulator entity. Unlike in the case of bacteria, other proteins would not be
present and consequently the overall characterization and modeling of the system would be
simpler. On the other hand, the regeneration of the pool of protons or other ions inside the
liposomes have to be refilled during the transmission process. The establishment and adjustment
of such an artificial system, based on different protein complexes, will be addressed in the future.

\section{Mathematical Modelling of the End-to-End MC System}\label{Sec:Analytic}
In this section, we develop an analytical model to characterize the chemical signal induced by the bacteria as a function of the applied optical signal.

\subsection{System Step Response to Illumination and Darkness}\label{Sec:StepResp}

Let $T^{\mathrm{symb}}$ denote the length of a symbol interval. In this paper, we consider OOK modulation where for binary zero, the LED is turned off for the entire symbol interval whereas for binary one, the LED is turned on from the beginning of the symbol interval until time $\alpha T^{\mathrm{symb}}$ and is turned off for the remaining time $(1-\alpha) T^{\mathrm{symb}}$ of the symbol interval, i.e., a rectangular pulse shape is adopted.
The motivation for adopting this specific pulse shape is to avoid a constant increase of proton concentration when pulses are transmitted consecutively, e.g., corresponding to transmission of consecutive binary ones for OOK.  To characterize this system, in the following, we develop an analytical parametric model for the step response of the system to  illumination and darkness.  This analytical model is based on a simplified physical system and heuristic modifications to account for the complexities of the considered biological~system.

\subsection{Mathematical Model under Simplifying Assumptions}
\label{sec:physical_model}
Analytical models to describe the proton release rate or proton concentration (or equivalently the pH) as a function of a given induced optical intensity have been developed in \cite{Hamid_Modulator,zifarelli2008buffered}. In particular, in \cite{Hamid_Modulator}, the photocycle of the bacteriorhodopsin was modeled as a Markov chain and the corresponding proton release rate was derived. Moreover, in \cite{zifarelli2008buffered}, the expected pH change in the proximity of a proton pumping cell was derived as a function of time. In this paper, we do not aim to fully model the considered biological system based on the laws of physics  since the system is too complex to lend itself to a tractable analysis. Nevertheless, to facilitate analytical insight, we make some simplifying  assumptions for the considered biological system which enable the derivation of a simple mathematical model. In the next subsection, we propose further modifications of this mathematical model to account for the peculiarities observed in the measurement data.  

A qualitative description of the bacterial system was provided in Section~\ref{Sec:Kinetic}. In the following, we make some assumptions to arrive at a first-order approximation for the response of the considered system to illumination and darkness. 

\begin{itemize}
\item[\textbf{A1:}] We assume that bacteria are uniformly distributed in the container and the light stimulus affects all bacteria in the same manner. This approximation is accurate since the bacteria solution is well-stirred.
\item[\textbf{A2:}] We assume that within each bacterial cell, there is an inexhaustible pool of protons which does not deplete despite pumping. Moreover, the pumps can only be in two states, fully on or fully off. These states are controlled by the light stimulus and are not affected by the concentration gradient across the cell membrane \cite{Hamid_Modulator}. This assumption is approximately valid since unlike passive transporters such as ion channels, proton pumps are active transporters which are able to pump  protons out of the cell membrane using the absorbed light energy \cite{Gadsby2009ion}.  Therefore, we assume that upon illumination, all pumps release protons into the channel with a fixed effective rate, denoted by $\rho$. 
\item[\textbf{A3:}] 
The bacterial cells take up protons in a passive manner such as via proton ion channels until an equilibrium value $\ceqdark$ is reached.
This uptake is proportional to the difference of the proton concentration, denoted by $c(t)$, and the equilibrium proton concentration $\ceqdark$ \cite{Maffeo2012modeling}.
\item[\textbf{A4:}]
The proton uptake depends on the illumination state.
In other words, the effective rate of proton uptake is $\beta_i(c(t)-\ceqdark),\,\,i\in\{0,1\}$, where $\beta_i$ is the uptake rate constant which is dependent on whether the illumination is on ($i=1$) or off ($i=0$).
This assumption is motivated by the complicated interactions between the buffer liquid and the machinery of the cells where, e.g., the proton pumping induces several ion currents which in turn affect the uptake of protons \cite{Gadsby2009ion,Maffeo2012modeling}.
\end{itemize}

Based on the above assumptions, the proton concentration for the considered system under darkness ($i=0$) and illumination ($i=1$) can be mathematically modeled by the following ordinary differential equation 
\begin{IEEEeqnarray}{rCl}
    \label{eq:physical_ode}
    \ddt c(t) &=& i \cdot\pumprate - \beta_i\cdot (c(t) - \ceqdark ), \quad i\in\{0,1\},
\end{IEEEeqnarray}
with concentration $c(t_0)$ at initial time $t_0$. For the above equation, there exists a steady state solution which can be found by setting $\ddt c(t)=0$ as
\begin{equation}
 c_i^{\infty} \triangleq  \lim\limits_{t\to\infty} c(t)  = \ceqdark + \frac{i\cdot\pumprate}{\beta_i}.
\end{equation}
By defining auxiliary variable $x(t) = c(t)-c(t_0)$, i.e., the difference between the concentration at time $t$ and the concentration at initial time $t_0$, \eqref{eq:physical_ode} simplifies to
\begin{IEEEeqnarray}{rCl}
    \label{eq:ODE}
    \frac{\mathrm{d}x(t)}{\mathrm{d}t} &=& -\frac{1}{\tau_i} \big(x(t) - (c_i^{\infty} - c(t_0)) \big),
\end{IEEEeqnarray}
where $\tau_i=1/\beta_i$ and the initial condition is $x(t_0)=0$. We can obtain the solution to \eqref{eq:ODE} as
\begin{equation}
    \label{eq:xonoff}
    x(t) = (c_i^{\infty} - c(t_0)) \cdot \big(1 - e^{-\frac{t-t_0}{\tau_i}}\big),\quad t\geq t_0.
\end{equation}
The above solution for the considered simplified system model serves as the basis for the analytical model which we propose in the following subsection. In particular, under both the illumination ($i=1$) and darkness ($i=0$) states, (\ref{eq:xonoff}) suggests that the proton concentration converges exponentially fast towards the corresponding equilibrium level. We note that the values of the time constant $\tau_i$ and the equilibrium level $c_i^{\infty}$ differ under illumination ($i=1$) and darkness ($i=0$). This observation is in line with the results reported in \cite{zifarelli2008buffered}. 

\begin{remk}
Note that the optical signal for the considered OOK modulation in Section~\ref{Sec:StepResp} consists of a sequence of illumination and darkness intervals.  Hence, (\ref{eq:xonoff}) can be used to derive the chemical response, i.e., the proton concentration, of the considered biological system to a sequence of OOK optically modulated symbols, cf. Section~\ref{sec:OOK}.  
\end{remk}

\subsection{Proposed Analytical Model}
\label{sec:analytical_model}

Although the measurement data collected from our testbed is consistent with the exponential behavior predicted in (\ref{eq:xonoff}) (see also Fig.~\ref{Fig:SingleShot}), it also reveals the existence of two phenomena that are not captured by the simplified model in (\ref{eq:xonoff}):

\textit{i)} \textbf{Slow drift:} From our measurement data, we observed that the proton concentration may exhibit a drift behaviour that was neither predicted in \cite{Hamid_Modulator,zifarelli2008buffered} nor by the simplified physical model in Section~\ref{sec:physical_model}. In particular, our measurements reveal a slow drift over relatively long time intervals (e.g., on the order of \SI{20}{\minute}) compared to the considered symbol interval duration, which is on the order of \SI{1}{\minute}. This effect can be attributed to a slow variation in the behavior of the bacteria over time, e.g., due to aging of the bacteria.  The concentration change from initial time $t_0$ to time $t$ due to this drift is denoted by $d(t)$ and is modeled by the following linear deterministic function 
\begin{equation} 
 d(t)=m^{\mathrm{d}}(t-t_0),
\end{equation}
where $m^{\mathrm{d}}$ is the slope of the drift.

\textit{ii)} \textbf{Fast random fluctuations:} There are additional fluctuations in the concentration $c(t)$ that are much faster than the signal and drift components. We model these fluctuations as noise denoted by $e(t)$. This noise may include diffusion (counting) noise, pH sensor circuitry noise, and the noise inherent to the biological machinery of the bacteria. Because of the law of large numbers, one analytical approximation that is often accurate when the overall noise is the result of the superposition of many independent noise sources is to model the overall noise as Gaussian noise. In fact, we observed that for the measurements collected from \textit{each bacterial culture}, the noise distribution can be well approximated as Gaussian, see Section~\ref{Sec:Results}. In particular, we model the noise as
\begin{equation}\label{Eq:NoisePDF} 
	e(t)\sim\mathcal{N}(0,\sigma^2),
\end{equation}
where $\mathcal{N}(\mu,\sigma^2)$ denotes a Gaussian random variable with mean $\mu$ and variance $\sigma^2$. However, the value of $\sigma^2$ may change from one bacterial culture to the next due to the inevitable randomness that is inherent to bacterial culturing.
 
To summarize, the proposed model for the proton concentration comprises the aforementioned three components, namely signal, drift, and noise, and is given by
\begin{equation}\label{Eq:Model}
	c(t) = c(t_0) + x(t) + d(t) + e(t),\quad t\geq t_0.
\end{equation}
Note that, conditioned on the transmitted symbols, components $x(t)$ and $d(t)$ are modeled as deterministic, whereas $e(t)$ is modeled as random. Moreover, the drift component $d(t)$ and the noise component $e(t)$ in (\ref{Eq:Model}) do not depend on the optical signal.  In the following subsection, we derive an explicit expression for $x(t)$ as a function of the transmitted OOK-modulated optical signal for the specific pulse shape introduced in Section~\ref{Sec:StepResp}.

\subsection{OOK Received Signal}\label{sec:OOK}

Let us assume a time-slotted communication where time is divided into different time slots of length $T^{\mathrm{symb}}$ and transmission starts at $t=t_0=0$. In particular, we transmit $K$ consecutive symbols where $s_k\in\{0,1\}$ denotes the $k$-th OOK symbol and refer to all $K$ symbols as a frame. Using (\ref{eq:xonoff}), $x(t)$ in the $k$-th symbol interval, i.e., $t\in[(k-1)T^{\mathrm{symb}},kT^{\mathrm{symb}})$, is obtained as
\begin{equation}
    \label{eq:ook_model}
    x(t) =
    \begin{cases}
        \Big(\cinf_0 - \bar{c}\big((k-1)T^{\mathrm{symb}}\big)\Big)\Big(1-e^{-\frac{t-(k-1)T^{\mathrm{symb}}}{\tau_0}}\Big), \\
        \hspace{3cm} \mathrm{if}\,\, s_k=0\,\,\text{and}\,\,  t\in[(k-1)T^{\mathrm{symb}},kT^{\mathrm{symb}}) \\
        \Big(\cinf_1 - \bar{c}\big((k-1)T^{\mathrm{symb}}\big)\Big)\Big(1-e^{-\frac{t-(k-1)T^{\mathrm{symb}}}{\tau_1}}\Big), \\
        \hspace{3cm} \mathrm{if}\,\, s_k=1\,\,\text{and}\,\,  t\in[(k-1)T^{\mathrm{symb}},(k-1+\alpha)T^{\mathrm{symb}}) \\
        \Big(\cinf_0 - \bar{c}\big((k-1+\alpha)T^{\mathrm{symb}}\big)\Big)\Big(1-e^{-\frac{t-(k-1+\alpha)T^{\mathrm{symb}}}{\tau_0}}\Big), \\
        \hspace{3cm} \mathrm{if}\,\, s_k=1\,\,\text{and}\,\,  t\in[(k-1+\alpha)T^{\mathrm{symb}},kT^{\mathrm{symb}})
    \end{cases}
\end{equation}
where $\bar{c}(t)=\mathbb{E}\{c(t)\}=c(t_0) + x(t) + d(t)$ is the deterministic component of the model in  (\ref{Eq:Model}) and $\mathbb{E}\{\cdot\}$ denotes expectation.

The analytical model in (\ref{Eq:Model}) and \eqref{eq:ook_model} is used to derive estimation and detection schemes proposed in Section~\ref{Sec:Comm}. Furthermore, we can employ this model to generate artificial channel realizations.
This is useful for the design and evaluation of transmission schemes, which might require a large number of independent channel realizations, the generation of which via experiments may be too time consuming.

\section{Communication System Design}\label{Sec:Comm}
In this section, we develop a parametric channel estimator and different detectors for recovery of the transmitted data from the measured received signal.

\subsection{Channel Estimation}
Knowledge of the channel parameters, collected in vector $\boldsymbol{\theta}=\left[\tau_0,\tau_1,c_0^{\infty},c_1^{\infty},m^{\mathrm{d}},\sigma^2,c(t_0)\right]$, is needed for efficient data detection. To this end, prior to data detection, we first estimate these parameters using a pilot-based estimation framework \cite{TCOM_MC_CSI,Adam_Channel}. In particular, we assume that the first $N,\,\,N<K,$ symbols in a frame are pilot symbols, denoted by $\mathbf{s}=[s_1,s_2,\dots,s_{N}]$, that are known at the receiver. Moreover, we assume that the receiver samples the received signal at time instances $t_n=(n-1)\Delta t,\,\,n=1,2,\dots$, where $\Delta t$ is the sampling interval\footnote{Without loss of generality, we assume $\Delta t$ is chosen such that  $T^{\mathrm{symb}}/\Delta t$ is an integer number.}. Furthermore, we employ the least square (LS) error as the performance criterion for parameter estimation.  The estimated parameters, denoted by $\hat{\boldsymbol{\theta}}=\left[\hat{\tau_0},\hat{\tau_1},\hat{c}_{0}^{\infty},\hat{c}_{1}^{\infty},\hat{m}^{\mathrm{d}},\hat{c}(t_0)\right]$, are given as follows
\begin{equation}\label{Eq:LSprob}
	\hat{\boldsymbol{\theta}} = 
    \underset{\bar{\boldsymbol{\theta}}}{\mathrm{arg\,min}}\,\, {g}{\left(\bar{\boldsymbol{\theta}}\right)}
\end{equation}
with
\begin{equation}\label{Eq:LSmetric}
    {g}{\left(\bar{\boldsymbol{\theta}}\right)} = \frac{1}{|\mathcal{T}_{\mathrm{tr}}|}
	\sum_{t_n\in\mathcal{T}_{\mathrm{tr}}} \,\,\left|c(t_n)-\bar{c}(t_n|\mathbf{s},\bar{\boldsymbol{\theta}})\right|^2,
\end{equation}
where $\bar{\boldsymbol{\theta}}=\left[\tau_0,\tau_1,c_0^{\infty},c_1^{\infty},m^{\mathrm{d}},c(t_0)\right]$, $\mathcal{T}_{\mathrm{tr}}=\{t_n|n=1,\dots,N T^{\mathrm{symb}}/\Delta t\}$ is the set of time instances used for training, and $\bar{c}(t_n|\mathbf{s},\bar{\boldsymbol{\theta}})$ is the deterministic component of the model in  (\ref{Eq:Model}) for a given pilot sequence $\mathbf{s}$ and channel parameters $\bar{\boldsymbol{\theta}}$. Here, $|\cdot|$ denotes the absolute value of a number or the cardinality of a set. The above problem can be solved numerically using algorithms such as the Newton-Raphson method to avoid having to perform an exhaustive search, see \cite{Adam_Channel} for a detailed treatment of a similar estimation problem. In fact, standard off-the-shelf solvers, such as the Matlab Curve Fitting Toolbox\texttrademark, can be used to find the solution of (\ref{Eq:LSprob}). Once $\hat{\boldsymbol{\theta}}$ is known, an estimate of the noise variance, denoted by $\hat{\sigma}^2$, can be obtained as 
\begin{equation}\label{Eq:Noise}
	\hat{\sigma}^2 = g\big(\hat{\boldsymbol{\theta}}\big).
\end{equation}

\subsection{Data Detection}
In the following, we first formulate the optimal detector which is based on the developed statistical model and assumes perfect channel estimation. This detector is complex and sensitive to the accuracy of the adopted model and the estimated channel parameters. To address these issues, we also propose a suboptimal low-complexity detector which is model free and hence does not  explicitly  rely on the estimated channel parameters.

\subsubsection{Optimal Detector}

To avoid the complexity of joint multiple-symbol detection, in the following, we focus on symbol-by-symbol detection. In particular, we consider a genie-aided ML detector that assumes perfect knowledge of the previously transmitted sequence in each symbol interval. In practice, the receiver employs the previously detected symbols instead of the genie symbols. Let $\mathbf{s}_k=[s_1,s_2,\dots,s_{k-1}]$ denote the  symbols transmitted before symbol interval $k$. The genie-aided ML detection
problem for the considered transmission scheme is given by
\begin{equation}\label{Eq:MLprob}
	\hat{s}_k = \underset{s_k\in\{0,1\}}{\mathrm{arg\,max}}\,\,\mathrm{Pr}\left\{c(t_n|s_k,\mathbf{s}_k)|_{\forall t_n\in\mathcal{T}_k}\right\}
	\overset{(a)}{=} \underset{s_k\in\{0,1\}}{\mathrm{arg\,min}}\,\, f\big(s_k,\mathbf{s}_k,\hat{\boldsymbol{\theta}}\big),
\end{equation}
where 
\begin{equation}\label{Eq:MLmetric}
f\big(s_k,\mathbf{s}_k,\hat{\boldsymbol{\theta}}\big) = \frac{1}{|\mathcal{T}_k|}\sum_{t_n\in\mathcal{T}_k} \,\,\big|c(t_n)-\bar{c}(t_n|s_k,\mathbf{s}_k,\hat{\boldsymbol{\theta}})\big|^2,
\end{equation}
where $\mathrm{Pr}\left\{\cdot\right\}$ denotes probability, $\mathcal{T}_k=\{t_n|n=(k-1)T^{\mathrm{symb}}/\Delta t+1,\dots,kT^{\mathrm{symb}}/\Delta t\}$, equality $(a)$ in (\ref{Eq:MLprob}) follows from the assumed Gaussian model, and $\bar{c}(t_n|s_k,\mathbf{s}_k,\hat{\boldsymbol{\theta}})$ is the deterministic component of the model in  (\ref{Eq:Model}) for given hypothesis symbol $s_k$, previously transmitted sequence $\mathbf{s}_k$, and estimated channel parameters~$\hat{\boldsymbol{\theta}}$.

\subsubsection{Suboptimal Low-Complexity Detector}
The optimal detector in (\ref{Eq:MLprob}) exploits the proposed underlying statistical channel model as well as the estimated channel parameters. Therefore, the ML detector is sensitive to the accuracy of the proposed model and the estimated channel parameters.   To cope with these issues, in the following, we propose a simple suboptimal detector which is motivated by the characteristics of the experimental data collected from our testbed. In particular, we observed from the measurements that the two dominant impairments are the short-term random fluctuation due to noise, $e(t)$, and the long-term drift, $d(t)$. Therefore, we employ a smoothing filter to mitigate the random fluctuations and a differential detector to eliminate the drift. The smoothed signal, denoted by $c^{\mathrm{sm}}(t_n)$, and the differentiated signal, denoted by $c^{\mathrm{df}}(t_n)$,  are given by
\begin{IEEEeqnarray}{ll}\label{Eq:Signal}
c^{\mathrm{sm}}(t_n) &= \frac{1}{T^{\mathrm{sm}}/\Delta t+1}\sum_{\tau\in\mathcal{T}_{n}^{\mathrm{sm}}} c(\tau) \quad \text{and}\\
c^{\mathrm{df}}(t_n) &= \frac{c^{\mathrm{sm}}(t_n+T^{\mathrm{df}})-c^{\mathrm{sm}}(t_n)}{T^{\mathrm{df}}},
\end{IEEEeqnarray}
respectively, where $\mathcal{T}^{\mathrm{sm}}_{n}=\{t_n|1+t_n/\Delta t\leq n \leq 1+(t_n+T^{\mathrm{df}})/\Delta t\}$, and $T^{\mathrm{sm}}$ and $T^{\mathrm{df}}$ are the lengths of the smoothing window and the differentiation window, respectively\footnote{Without loss of generality, $T^{\mathrm{sm}}$ and $T^{\mathrm{df}}$ are chosen such that $T^{\mathrm{sm}}/\Delta t$ and $T^{\mathrm{df}}/\Delta t$ are integer numbers, respectively.}. Thereby, we define $q[k]=\underset{t_n\in\mathcal{T}_k}{\max} \,\,c^{\mathrm{df}}(t_n)$ as the detection metric which enables us to employ the following simple threshold detector to recover the data 
\begin{equation}\label{Eq:ThrDet}
\hat{s}_k = \begin{cases}
1, \quad &q[k] \geq \eta\\
0, &\mathrm{otherwise},
\end{cases}
\end{equation}
where $\eta$ is the detection threshold. Note that the above detector does not require knowledge of the channel parameters. Therefore, we can use the pilot sequence to directly estimate the detection threshold. To this end, we propose the following detection threshold
\begin{equation}\label{Eq:Thr}
\eta =  \frac{\gamma}{|\mathcal{S}_0|}\sum_{k\in\mathcal{S}_0} q[k]+ \frac{1-\gamma}{|\mathcal{S}_1|}\sum_{k\in\mathcal{S}_1} q[k],
\end{equation}
where $\mathcal{S}_0$ and $\mathcal{S}_1$ contain the indices of pilot symbols which are zero and one, respectively, and $\gamma\in(0,1)$ is a design parameter.

\subsection{Adaptive Transmission}\label{Sec:Adaptive}

Recall that in the proposed transmission strategy, the parameters of the system are estimated via pilot symbols inserted at the beginning of each frame and are used for detection of the data symbols in the remaining portion of the frame. A challenge that arises with such a transmission scheme is that the  parameters estimated at the beginning of the frame gradually become  outdated during the frame as time progresses. In the considered biological system, there are two main reasons for this behavior. First, as discussed in Section~\ref{Sec:Variability}, biological systems are expected to undergo variations over time, e.g. due to changes in the dynamics of the bacteria caused by their aging. From a mathematical point of view, such a time-variant behavior implies that the parameters of the considered model vary over time. Second, the actual biological system is very complex and the proposed statistical model is a first-order approximation. Therefore, even if the system was time-invariant, the signal predicted by our model would diverge from the measurement signal due to the model mismatch. To address this issue, we propose the following adaptive transmission schemes.

\textit{i)} \textbf{Short Frame--Pilot-based Estimation:} A straightforward approach to tackle this problem is to choose the frame size sufficiently small such that the aforementioned variations are limited within the frame. Thereby, the model parameters are re-estimated at the beginning of the subsequent frame before they are significantly outdated. Assuming a fixed pilot sequence length to ensure a given estimation quality, the drawback of this scheme is that the overhead for channel estimation increases as the frame length decreases.

\textit{ii)} \textbf{Long Frame--Data-aided Estimation:} Similar to the previous scheme, here we first estimate the model parameters based on the pilot symbols at the beginning of the frame. Since the parameters become outdated as time progresses, we re-estimate the model parameters with a certain frequency within the frame. However, in this case, we employ the previously detected data symbols for estimation instead of pilot symbols. This adaptive scheme has a lower overhead compared to the previous scheme; however, the quality of parameter estimation deteriorates if the data symbols used for estimation are incorrectly detected. 

\section{Experimental Verification}\label{Sec:Results}
In this section, we present and analyze experimental data obtained with the proposed optical-to-chemical signal conversion interface. In particular, we first verify the analytical model derived in Section~\ref{Sec:Analytic}. Subsequently, we evaluate the performance of the estimation and detection schemes developed in Section~\ref{Sec:Comm}. Note that for all measurements, data transmission is preceded by a period of dark adaptation of \SI{30}{\minute}. The sampling rate of the pH signal is \SI{1}{\hertz}, i.e., one sample/second.  We employ the least square error criterion of the Matlab Curve Fitting Toolbox\texttrademark~ to estimate the model parameters $\boldsymbol{\theta}$, cf. (\ref{Eq:LSprob}).

\subsection{Verification of the Proposed Analytical Model}	

\begin{figure}
\center
\resizebox{0.8\linewidth}{!}{\psfragfig{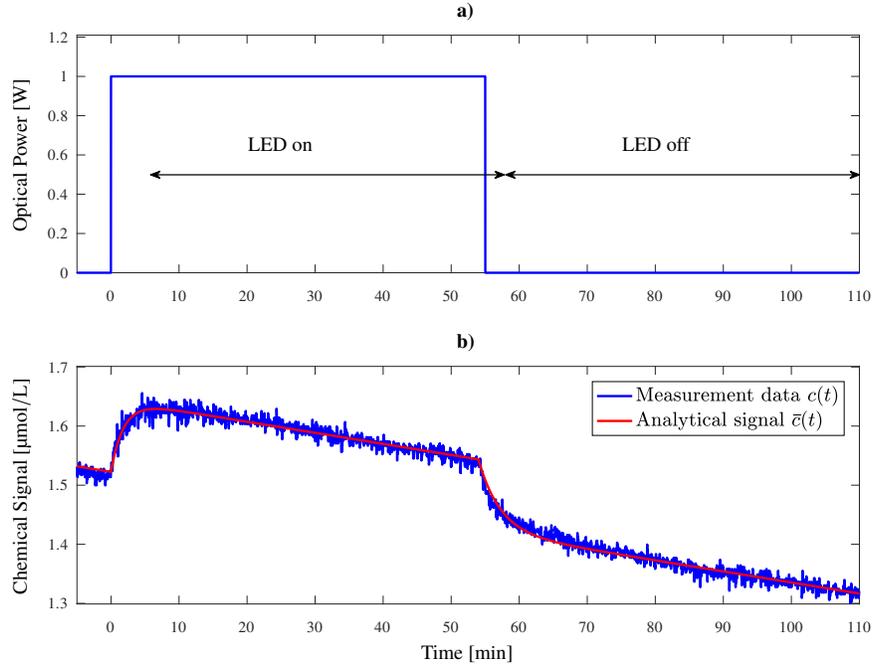}} 
    \caption{%
       a) Optical signal; b) Measured proton concentration $c(t)$ and analytical signal $\bar{c}(t)$ vs. time.
    }\label{Fig:SingleShot}
\end{figure}

In the following, the accuracy of the model developed in Section~\ref{Sec:Analytic} is investigated. First, in order to study the effect of the drift, we consider long illumination and darkness intervals with durations of \SI{55}{\minute}, respectively. In Fig.~\ref{Fig:SingleShot}a, we show the optical signal and in Fig.~\ref{Fig:SingleShot}b, we show the corresponding measured concentration $c(t)$ and the analytical signal $\bar{c}(t)$ for one cell culture  vs. time. The parameters of the proposed model are found as $c_0^{\infty}=\SI{1.53}{\micro\mol/\liter}$ (pH of $5.82$), $c_1^{\infty}=\SI{1.65}{\micro\mol/\liter}$ (pH of $5.78$), $\tau_0=\SI{3.19}{\minute}$, $\tau_1=\SI{1.85}{\minute}$, and $m^{\mathrm{d}}=\SI{-0.0019}{\micro\mol/\liter/\minute}$. As expected, the concentration level increases during illumination and decreases during darkness; hence, the optical signal is successfully converted to a chemical signal.  From the measured concentration shown in Fig.~\ref{Fig:SingleShot}b, we observe a baseline drift during both the illumination and darkness intervals. Overall, the proposed analytical signal is in very good agreement with the measurement~data. 
Thereby, we obtain for the sum of squared errors a value of \SI{0.4596}{(\micro\mol/\liter)^2} for a total of \num{6918} samples.

\begin{figure}
\center
    \resizebox{0.8\linewidth}{!}{\psfragfig{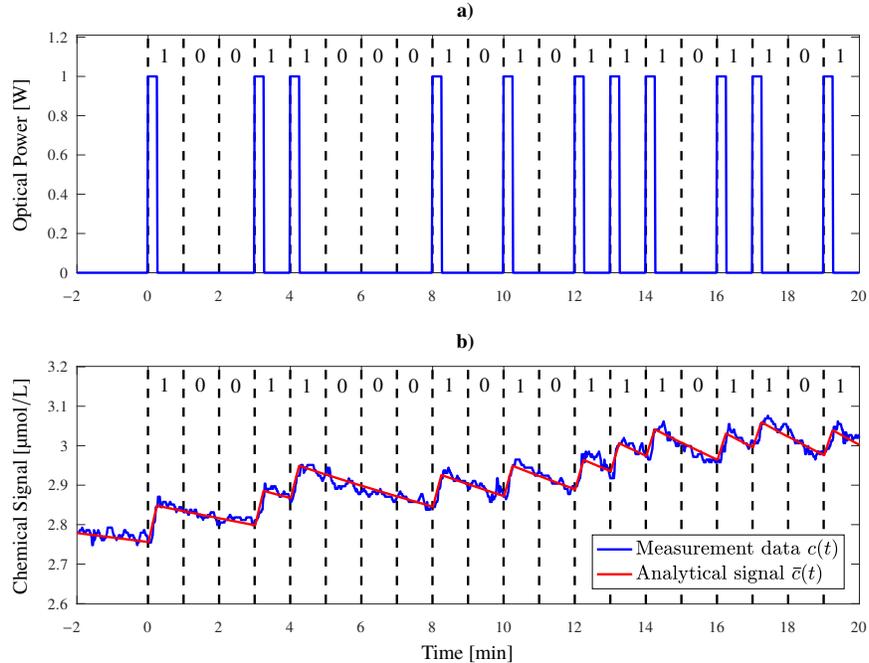}} 
    \caption{%
       a) Optical  signal corresponding to symbol sequence $[10011000101011101101]$; b) Measured proton concentration $c(t)$ and analytical signal $\bar{c}(t)$ vs. time. The dashed lines represent the start of a new symbol interval.
    }\label{Fig:MultipleShot}
\end{figure}

In Fig.~\ref{Fig:MultipleShot}a, we show the optical signal corresponding to a sequence of $20$ symbols, $[10011000101\\011101101]$, with $T^{\mathrm{symb}}=1$ min and $\alpha=0.25$, and in Fig.~\ref{Fig:MultipleShot}b, we show the corresponding measured concentration $c(t)$ and the analytical signal $\bar{c}(t)$  vs. time.  The parameters of the proposed model are found as $c_0^{\infty}=\SI{2.83}{\micro\mol/\liter}$  (pH of $5.55$), $c_1^{\infty}=\SI{5.77}{\micro\mol/\liter}$ (pH of $5.24$), $\tau_0=\SI{6.41}{\minute}$, $\tau_1=\SI{8.40}{\minute}$, and $m^{\mathrm{d}}=\SI{-0.0039}{\micro\mol/\liter/\minute}$. The values of the model parameters are different from those obtained for the measurements shown in Fig.~\ref{Fig:SingleShot} since the measurements shown in Fig.~\ref{Fig:MultipleShot} were gathered from different bacterial cultures. Again, we observe from Fig.~\ref{Fig:MultipleShot}a that the proposed analytical model explains the measurement data well even if multiple symbols are transmitted.
Thereby, we obtain for the sum of squared errors a value of \SI{0.2189}{(\micro\mol/\liter)^2} for a total of \num{2110} samples.

\begin{figure}
\center
\resizebox{0.8\linewidth}{!}{\psfragfig{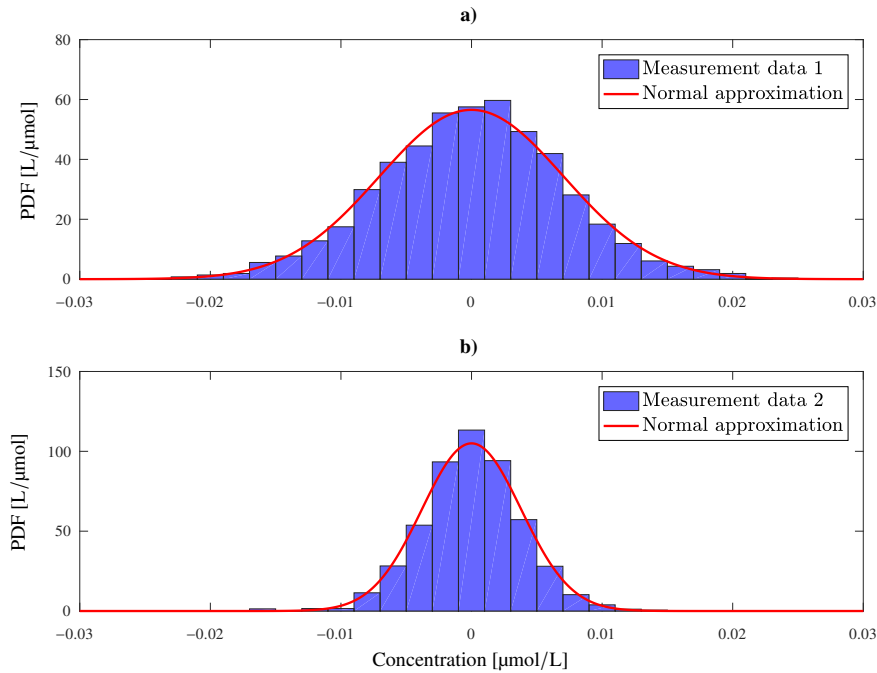}} 
\caption{Normalized histograms of the random fluctuation component $e(t)$ of two different measurements and the corresponding fitted normal probability density functions (PDFs).
    }\label{Fig:NoiseStat}
\end{figure}

Next, we study the statistical characteristics of the random fluctuation component $e(t)$. By subtracting the measured concentration from the fitted analytical model, we obtain the noise component as $e(t) = c(t)-\bar{c}(t)$. We show the results for two sets of measurements: \textit{i)} The LED is turned on for \SI{8}{\minute} and then turned off for \SI{16}{\minute}. \textit{ii)} The LED is turned on for \SI{20}{\minute} and then turned off for \SI{100}{\minute}. Note that these  two measurements are taken with different bacterial cultures. Fig.~\ref{Fig:NoiseStat} shows the histograms of $e(t)$ for the two measurements and the corresponding normal distribution that has the same mean $\mu$ and variance $\sigma^2$. For the measurement in Fig.~\ref{Fig:NoiseStat}a), we obtain $(\mu,\sigma) = (\num{5.54e-10},\num{0.0071})$~\si{\micro\mol/\liter} whereas for the measurement in Fig.~\ref{Fig:NoiseStat} b), we obtain $(\mu,\sigma) = (\num{2.91e-10},\num{0.0038})$~\si{\micro\mol/\liter}. As can be seen from this figure, the means for both measurements are small and close to zero. Moreover, the normal distribution approximates the actual noise histogram reasonably well which confirms that for \textit{a given bacterial culture}, the Gaussian statistical model proposed in (\ref{Eq:NoisePDF}) is a good approximation. However, as can be seen from Fig.~\ref{Fig:NoiseStat},  the noise variance for these two measurements differs noticeably which is due to the inevitable randomness that is inherent to the bacterial culturing process.

\subsection{Evaluation of the Developed Transmission Strategies}

In the following, we study the performance of the estimation and detection schemes developed in Section~\ref{Sec:Comm}. Our data set is the received signal corresponding to $600$ consecutively transmitted symbols; however, for clarity of presentation, we show the results only for the first $120$ symbols. Moreover, we assume $T^{\mathrm{symb}}=\SI{1}{\minute}$, $\alpha=0.25$, and $N=10$ pilot symbols. For the short-frame transmission, denoted by ``pilot-based", we adopt a frame length of $K=40$ symbols where the model parameters are estimated using the pilot symbols and are used for detection for the remaining part of the frame. For the long-frame transmission, denoted by ``data-aided", we adopt a frame length of $K=120$ symbols where the parameters are re-estimated within the frame with a frequency of $10$ symbols based on the $20$ previously detected data symbols. Therefore, the pilot-based and data-aided schemes have overheads of $\frac{1}{4}$ and $\frac{1}{12}$, respectively.

\begin{figure}
\center   
    \resizebox{1\linewidth}{!}{\psfragfig{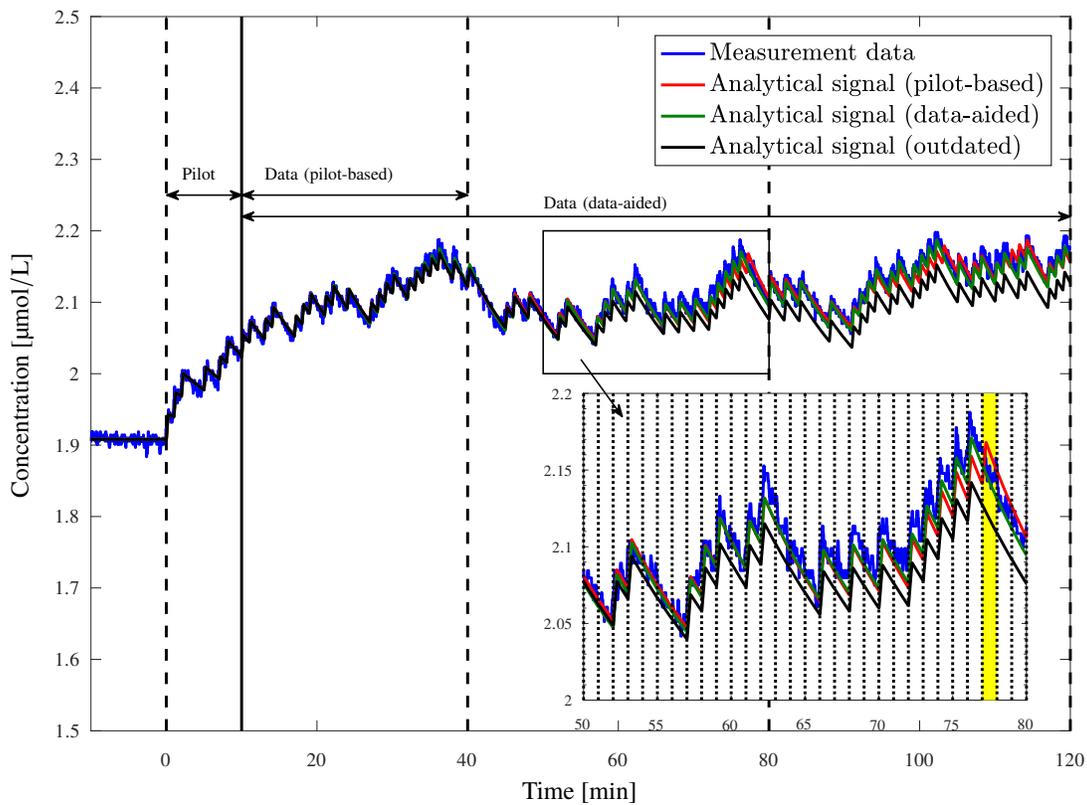}}
        \vspace*{-15mm}
    \caption{Measured proton concentration and corresponding analytical signals vs. time. The vertical dashed lines indicate the beginning of a new frame for the short-frame transmission scheme (denoted by ``pilot-based") whereas the vertical solid lines indicate the re-estimation within the frame for the long-frame transmission scheme (denoted by ``data-aided"). The vertical dotted lines represent the beginning of a new symbol interval.
    }\label{Fig:PilotDataEst}
\end{figure}

In Fig.~\ref{Fig:PilotDataEst}, we show the measured concentration and the analytical signal for the aforementioned transmission schemes vs. time. To illustrate that the estimated parameters gradually become outdated, we also show the analytical signal obtained using the  parameters estimated based on the pilot symbols in $t\in[0, 10]$~\si{\minute} and the \textit{actual} transmitted data, and denote the corresponding curve by ``outdated". In this case, we used the actual data to focus solely on the effect of the outdated parameters. As can be seen from Fig.~\ref{Fig:PilotDataEst}, the analytical model based on the outdated signal starts to diverge from the measurement data after $57$ min (corresponding to the $58$-th transmitted symbol). This underlines the need for the adaptive transmission schemes introduced in Section~\ref{Sec:Adaptive} and explains why we chose the frame length for short-frame transmission as $40$ symbols. Here, we use the ML detector in (\ref{Eq:MLprob}) for both the pilot-based and data-aided schemes. Note that unlike the outdated analytical signal, the pilot-based and data-aided schemes are impaired by incorrect data detection.  This is evident from the comparison of the curves for the outdated and pilot-based analytical signals in the zoomed plot in Fig.~\ref{Fig:PilotDataEst}. In this time interval, i.e., $t\in[10,40]$~\si{\minute}, both the outdated and the pilot-based analytical signals employ the same channel parameters, which were estimated from the pilot symbols in $t\in[0,10]$~\si{\minute}.
For the pilot-based scheme, an error occurs for symbol $78$, highlighted by a yellow vertical bar, because of a mismatch between the estimated model and the actual data. Compared to the pilot-based scheme, for the considered example, the data-aided scheme can follow the measurement data longer due to more frequent re-estimation of the channel parameters. However, as discussed in Section~\ref{Sec:Adaptive}, its performance is affected by imperfect data detection in the long run.
In summary, from Fig.~\ref{Fig:PilotDataEst}, we conclude that the proposed adaptive schemes can fairly well follow the measurement data. However, it is important to emphasize that their performances heavily depend on the choice of parameters such as the pilot length, frame length, and re-estimation frequency.

\begin{figure}
\center 
    \resizebox{1\linewidth}{!}{\psfragfig{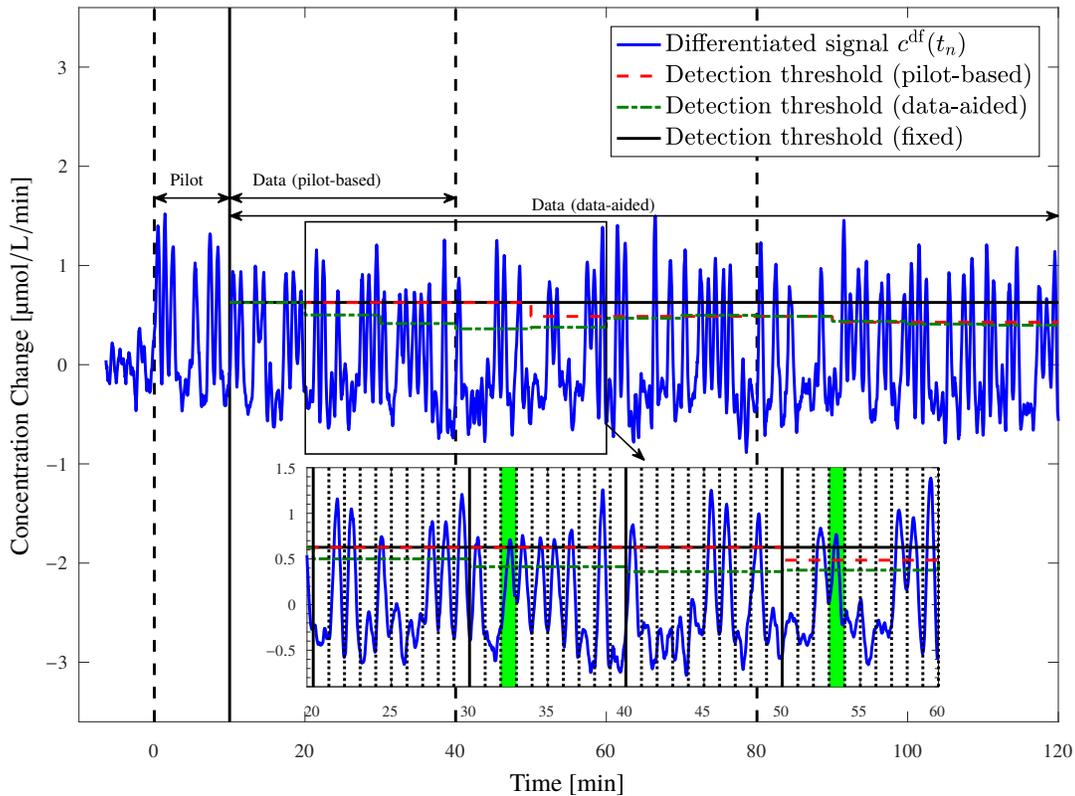}}
        \vspace*{-15mm}
    \caption{
        The differentiated signal $c^{\mathrm{df}}(t_n)$ used for data detection with the proposed suboptimal detector and the corresponding detection thresholds vs. time. The vertical dashed lines indicate the beginning of a new frame for the pilot-based scheme whereas the vertical solid lines indicate the re-estimation within the frame for the data-aided scheme. The vertical dotted lines represent the beginning of a new symbol interval. 
    }\label{Fig:SubOptimalDet}
\end{figure}

Next, we evaluate the performance of the proposed suboptimal low-complexity detector in (\ref{Eq:ThrDet}). The lengths of the smoothing and differentiation windows are chosen as $T^{\mathrm{sm}}=30$~s and $T^{\mathrm{df}}=20$~s, respectively. Moreover, we consider three schemes to determine the detection threshold in (\ref{Eq:Thr}) using $\gamma=0.5$. For the first scheme, referred to as ``fixed'', the detection threshold is computed only once based on the pilot symbols in $t\in[0,10]$~\si{\minute}. For the second scheme, referred to as ``pilot-based'', we adopt a short frame length of $40$ symbols and the detection threshold is updated in each frame based on the first $10$ pilot symbols. Finally, for the third scheme, referred to as ``data-aided'', we employ long frame lengths, and re-estimate the detection threshold within the frame with a frequency of $10$ symbols based on the $20$ previously detected symbols. In Fig.~\ref{Fig:SubOptimalDet}, we show the differentiated signal $c^{\mathrm{df}}(t_n)$ vs. time  which is used for data detection, cf. (\ref{Eq:ThrDet}). The positive peaks in the differentiated signal, which result from illumination when transmitting a binary ``1'', are very pronounced and substantially exceed the noise level. Compared to the original concentration, the differentiated signal $c^{\mathrm{df}}(t_n)$ exhibits much less variations over time which implies that its detection performance is more robust to channel variation than the ML detector in (\ref{Eq:MLprob}). Nevertheless, choosing a proper decision threshold is still challenging since the peak values do vary over time. As can be seen from Fig.~\ref{Fig:SubOptimalDet}, the proposed adaptive schemes are able to successfully track the channel variations by adapting their detection thresholds. For instance, since the peak values are relatively higher for the pilots symbols in $t\in[0,10]$~\si{\minute}, the corresponding fixed threshold is chosen large. Thereby, the fixed threshold scheme is able to successfully recover for example symbols $33$ and $54$ (highlighted by a green bar) \textit{only by a small margin}. On the contrary, the proposed pilot-based scheme (for symbol $54$) and data-aided scheme (for symbols $33$ and $54$) are able to successfully recover the data by a larger margin. 

Finally, we report the bit error ratio (BER) values corresponding to the $120$ symbols shown in Fig.~\ref{Fig:PilotDataEst} and Fig.~\ref{Fig:SubOptimalDet} as well as the entire  $600$ symbols used in this measurement. Note that the ML scheme was developed assuming that the underlying statistical model and the estimated parameters are perfectly known.  However, for the considered pilot-based and data-aided ML schemes, there is always a residual estimation error. Therefore, we also include results for a benchmark scheme, which we refer to as ``genie-aided ML", where the receiver re-estimates the channel parameters (similar to the pilot-based scheme but using all symbols in one frame as pilots) assuming perfect knowledge of the data symbols and then employs the estimated parameters for data detection. As can be seen from Table~I, the BER of the genie-aided ML scheme is zero for both considered numbers of symbols whereas due to the residual parameter estimation errors,  the pilot-based and data-aided ML schemes lead to higher BERs. Note that the different BERs of the pilot-based and data-aided schemes for $120$- and $600$-symbol sequences are due to the limited measurement data. Unlike the pilot-based and data-aided ML schemes, the proposed suboptimal scheme yields a non-zero BER only for a fixed decision threshold and transmission of \num{600} symbols. The reason for this behavior is that  the main source for detection errors for the proposed detection schemes is the low quality of the parameter estimation which is due to the time-varying nature of the considered biological system. Hereby, the proposed suboptimal schemes yield better detection performance since their operation is less sensitive to estimation errors than that of the proposed ML schemes.

\begin{table}
\label{Table:BER}
\caption{BERs of the Proposed Detection Schemes for the Considered Measurement Data. \vspace{-0.2cm}} 
\begin{center}
\scalebox{0.8}
{
\begin{tabular}{|| c | c | c | c ||}
  \hline 
  \multicolumn{4}{||c||}{\textbf{ML Schemes}}\\ \hline \hline 
 & Genie-aided&Pilot-based&Data-aided \\ \hline
 BER ($120$ Symbols) & $0$ & $3/90$ & $0/110$ \\ \hline
 BER ($600$ Symbols) & $0$ & $8/450$ &  $10/590$\\ \hline 
\end{tabular}
\begin{tabular}{|| c | c | c | c ||}
  \hline 
  \multicolumn{4}{||c||}{\textbf{Suboptimal Schemes}}\\ \hline \hline 
 & Fixed &Pilot-based&Data-aided \\ \hline
 BER ($120$ Symbols) & $0/110$ & $0/90$ & $0/110$ \\ \hline
 BER ($600$ Symbols) & $3/590$ & $0/450$ &  $0/590$\\ \hline 
\end{tabular}
}
\end{center}\vspace{-0.6cm}
\end{table}

\section{Conclusions and Future Work}\label{Sec:Conclusion}

In this paper, we introduced a biological microscale modulator based on \textit{E.~coli} bacteria that express the light-driven proton pump gloeorhodopsin and, in response to external light stimuli, can locally change their surrounding pH level by pumping protons into the channel. We provided an analytical model for characterization of the induced chemical signal as a function of the applied optical signal. We further derived estimation and detection schemes to recover the transmitted data from the induced chemical signal measured by a pH sensor which served as receiver. From our measurement data, we observed that the characteristics of the induced chemical signal vary over time. To cope with this issue, we developed adaptive transmission schemes which are able to follow the variations of the MC system. The proposed analytical model was shown to be in very good agreement with measurement data for a sequence of transmitted symbols. Furthermore, it was shown that the proposed setup is able to successfully convert an optical signal representing a sequence of binary symbols into a chemical signal with a bit rate of \SI{1}{bit/\minute} and recover the transmitted data from the chemical signal using the proposed estimation and detection schemes. We note that the high data rate of at least \SI{1}{bit/\minute} achieved by our testbed  is a big step forward compared to existing biological testbeds (e.g., the data rate of the system in \cite{Krishnaswamy_Time_2013} is approximately \SI{1}{bit/\hour}). In future work, we plan to replace the pH sensor by a bacterial receiver, e.g., bacterial cells expressing a pH-sensitive green fluorescent protein (GFP). Having both an optical-to-chemical converter as transmitter and a chemical-to-optical converter as receiver, we can set up a full MC system at microscale that can be efficiently controlled and read out at macroscale.



\bibliographystyle{IEEEtran}
\bibliography{tnbs.bbl}

\end{document}